\documentclass{elsart}

\usepackage{graphicx}

\usepackage{amssymb}

\begin{document}

\begin{frontmatter}

\title{Anisotropic thermal expansion and uniaxial pressure dependence of superconducting and magnetic transitions in ErNi$_2$B$_2$C.}

\author[AL]{S.L. Bud'ko},
\author[Oxy]{G.M. Schmiedeshoff},
\author[AL]{P.C. Canfield}
\address[AL]{Ames Laboratory US DOE and Department of Physics and Astronomy, Iowa State University, Ames, IA 50011, USA}
\address[Oxy]{Department of Physics, Occidental College, Los Angeles, CA 90041, USA}

\begin{abstract}
We present anisotropic thermal expansion measurements on single crystalline ErNi$_2$B$_2$C. All three,
superconducting, antiferromagnetic and weak ferromagnetic phase transitions are unambiguously distinguished in the
data. Anisotropic uniaxial pressure dependencies of the transitions are estimated based on the Ehrenfest relation,
leading to a conclusion, in particular, that weak ferromagnetic state will be suppressed by small, order of
several kbar, hydrostatic pressure.
\end{abstract}

\end{frontmatter}

Quaternary borocarbides, in particular the members of the $R$Ni$_2$B$_2$C ($R$ = Gd-Lu, Y) series of compounds,
exhibit a plethora of complex phenomena: co-existence of local moment magnetism and superconductivity,
non-locality and flux line lattice transitions, heavy fermion physics and intricate metamagnetism
\cite{can98a,mul01a,mul02a,bud06a}. ErNi$_2$B$_2$C superconducts below $\sim 10 $ K \cite{cav94a,eis94a}, below
$T_N \approx 6$ K superconductivity coexists with an incommensurate antiferromagnetic state with the ordering wave
vector \textbf{q} $\approx 0.55$ \textbf{a}$^*$ \cite{zar95a,sin95a,cho95a}, and then, on further cooling, a weak
ferromagnetic component develops below $T_{WFM} \approx 2.3$ K \cite{can96a,cho01a,kaw02a}.

Despite a large number of publications on physical properties of ErNi$_2$B$_2$C, few of them address temperature
dependent structural changes in the material. The evolution of the lattice parameters between 360 K and 90 K, as
measured by single crystal X-ray diffraction, was reported in Ref. \cite{jae02a}.  A magnetoelastic
tetragonal-to-orthorhombic structural distortion was observed \cite{det97a} below $T_N$ by synchrotron X-ray
scattering. A $\lambda$-type anomaly in thermal expansion at $T_N$ was reported in STM measurements \cite{cre06a}.
And finally, a low temperature $H - T$ phase diagram was probed \cite{doe02a} using thermal expansion (in an
applied field) and magnetostriction measurements.

In this publication we report high resolution, anisotropic ($c$-axis and $a$-axis) thermal expansion (TE) and
longitudinal, low temperature, magnetostriction (MS) measurements for $H \| a$ on  single crystalline
ErNi$_2$B$_2$C. These measurements were performed with the intent to study the anomalies in TE associated with all
three salient transition temperatures (to the best of our knowledge, for two of them, $T_c$ and $T_{WFM}$, no TE
anomalies were reported so far) and to evaluate the uniaxial pressure derivatives of the three transition
temperatures using the the thermodynamic Ehrenfest relation.

Plate-like single crystals of ErNi$_2$B$_2$C with the $c$-axis perpendicular to the plates were grown by a Ni$_2$B
high temperature flux method (see Refs. \cite{can98a,cho95a,mxu94a,can01a} for more details). For the TE/MS
measurements the ErNi$_2$B$_2$C crystal was shaped into a nearly rectangular bar, with two pairs of parallel
surfaces, so that the measurements were performed along [100] ($L = 2.23$ mm) and [001] ($L = 1.42$ mm)
directions. After shaping, the sample was annealed in dynamic vacuum ($10^{-5}-10^{-6}$ Torr) at 950$^\circ$ C for
$\sim 100$ hours \cite{mia02a}. Thermal expansion and magnetostriction were measured using a capacitive
dilatometer constructed of OFHC copper; a detailed description of the dilatometer will appear elsewhere
\cite{sch06a}. The dilatometer was mounted in a Quantum Design PPMS-14 instrument and was operated over a
temperature range of 1.8 to 305 K and in magnetic fields up to 140 kOe. The same set-up was used in our recent
work on YNi$_2$B$_2$C \cite{bud06b}. Heat capacity was measured using a heat capacity option with $^3$He insert of
a PPMS instrument. DC magnetization was measured using a Quantum Design MPMS-7 SQUID magnetometer.

\begin{figure}
    \centering
    \includegraphics[height=70mm]{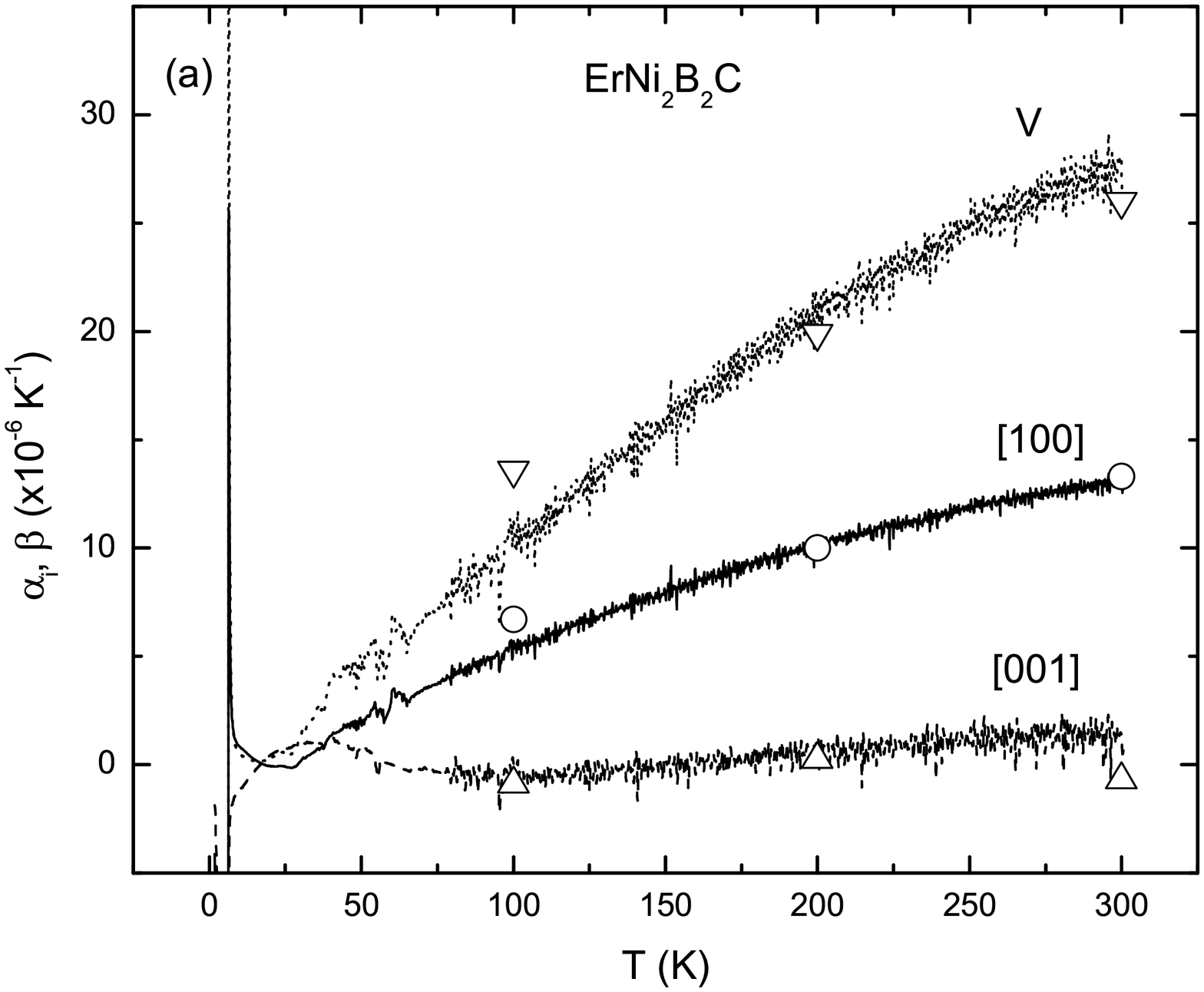}
    \includegraphics[height=70mm]{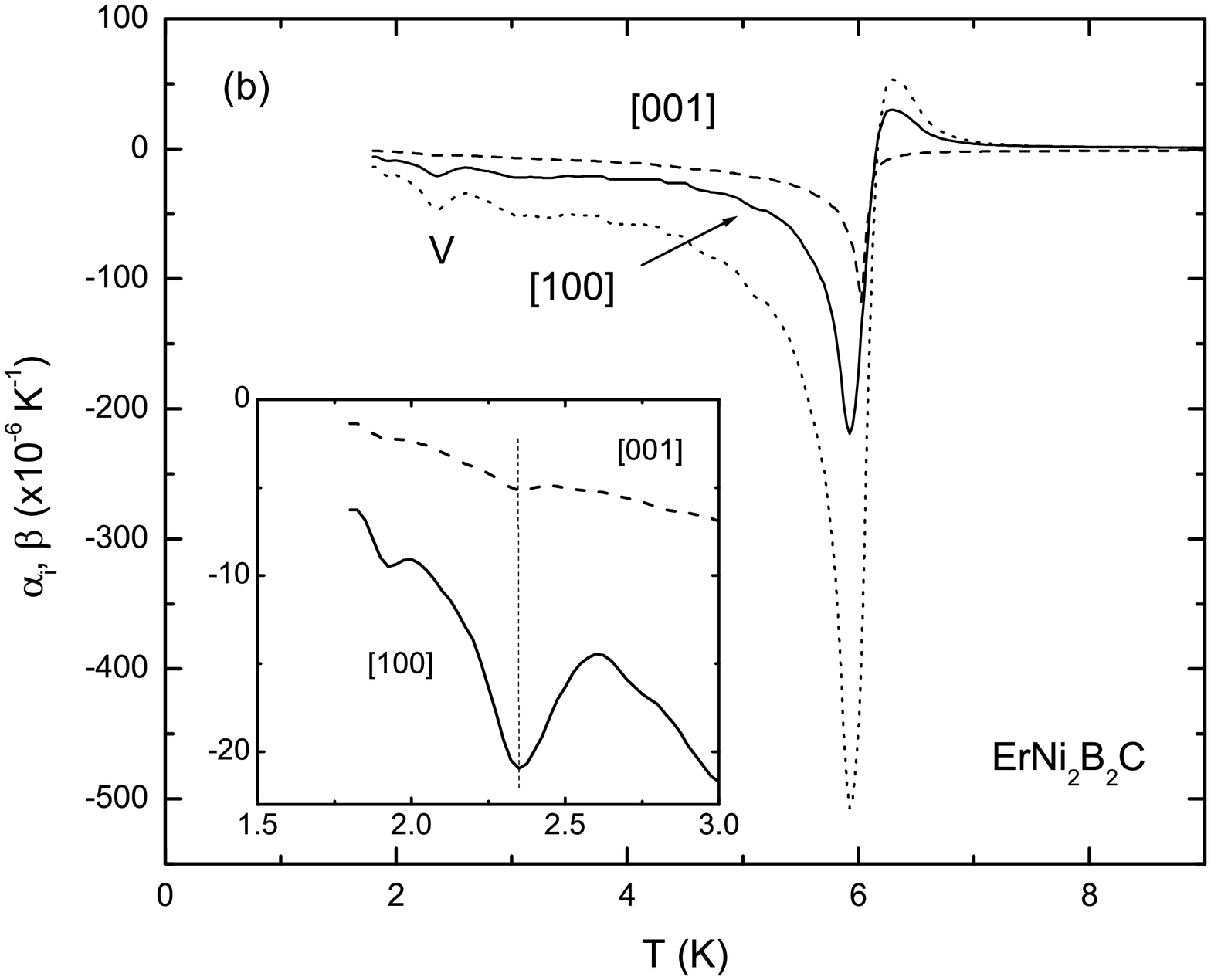}
    \includegraphics[height=70mm]{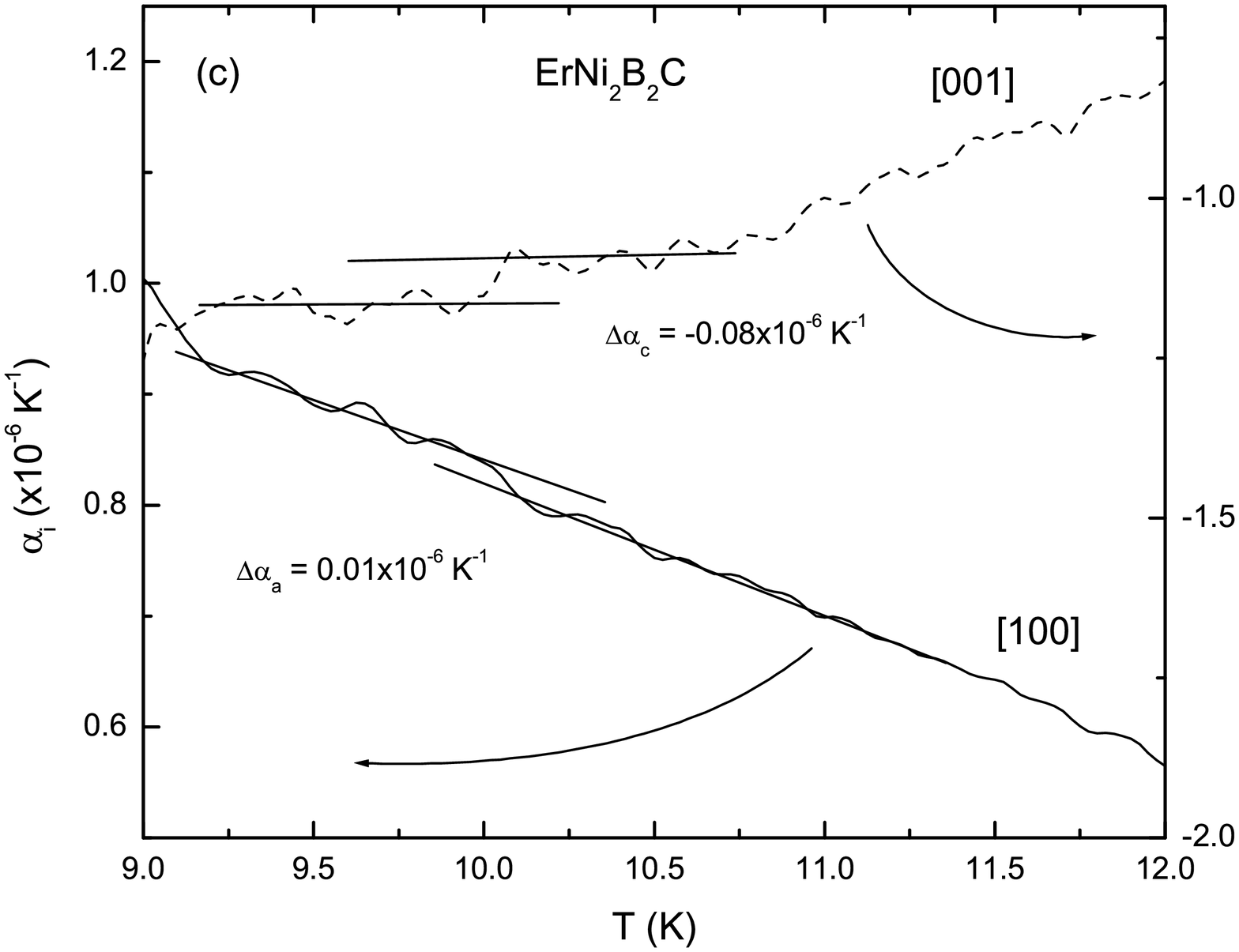}
    \caption{The temperature-dependent linear and volume thermal expansion coefficients of ErNi$_2$B$_2$C.
    Open symbols in (a) are data from Ref. \cite{jae02a}. (b) low temperature thermal expansion,
    inset - enlarged view near the transition to weak ferromagnetic phase, $T_{WFM}$ is marked
    with a vertical line. (c) linear TE coefficients near superconducting transition.} \label{F1}
\end{figure}

\begin{figure}
    \centering
    \includegraphics[height=100mm]{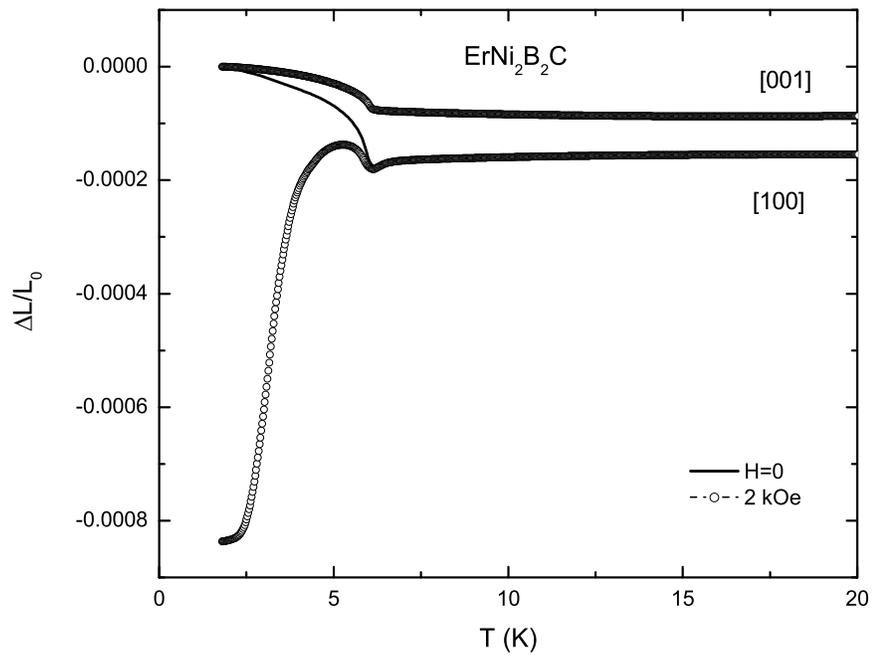}
    \caption{Low temperature longitudinal dilation of ErNi$_2$B$_2$C in zero and 2 kOe applied field
    plotted relatively to the $T = 1.8$ K, $H = 0$ ($L_0$) value. For $[001]$ the $H = 0$ and $H = 2$ kOe lines are
    indistinguishable on the scale of the graph.} \label{F2}
\end{figure}

\begin{figure}
    \centering
    \includegraphics[height=100mm]{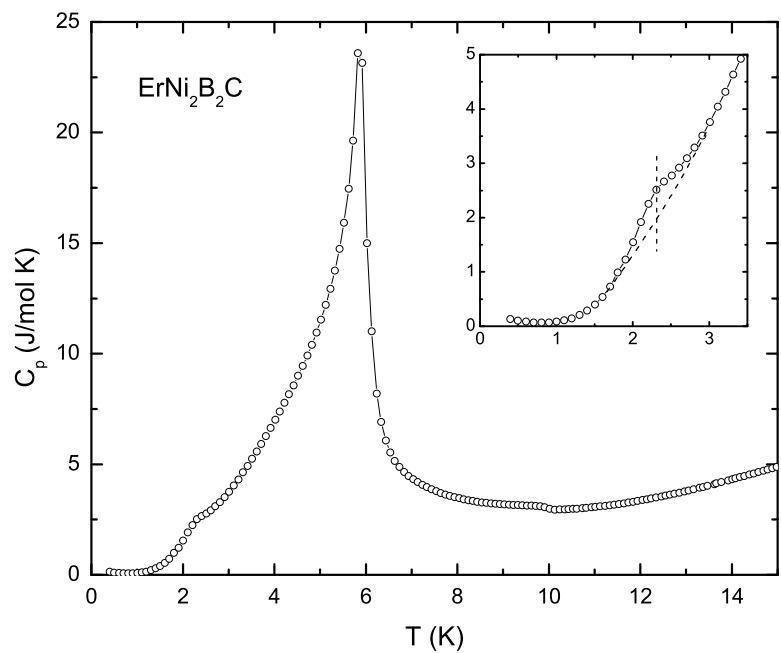}
    \caption{Temperature-dependent heat capacity of ErNi$_2$B$_2$C. Inset: enlarged region near $T_{WFM}$ with
    a sketch of an estimate of $\Delta C_p$ at $T_{WFM}$.} \label{F3}
\end{figure}

The temperature-dependent linear ($\alpha_a$, $\alpha_c$) and volume ($\beta = 2 \alpha_a + \alpha_c$) thermal
expansion coefficients are shown in Fig. \ref{F1}. Data for $T \geq 100$ K are in a good agreement with the
previously reported results \cite{jae02a} (shown as open symbols in Fig. \ref{F1}(a)). Clear anomalies are
detected at $T_N$ and $T_{WFM}$ (Fig. \ref{F1}(b)). The size of the (inverse) $\lambda$-like anomaly in $\alpha_c$
at $T_N$ is similar to the value reported in Ref. \cite{cre06a}. On cooling down from just above $T_N$ the
$\alpha_a$ initially increases and then goes into sharp, deep minimum. Such behavior (as opposed to some sort of
$\lambda$-like anomaly) may be associated with possible formation of the antiferromagnetic domains (twins) in the
crystal in zero or small fields \cite{vin05a,blu06a} due to orthorhombic structural distortion below $T_N$
\cite{det97a}. The complexity of in-plane thermal expansion of ErNi$_2$B$_2$C near $T_N$ is illustrated in Fig.
\ref{F2}: whereas for $H \| c$ an application of 2 kOe field does not change the temperature-dependent dilation
(and the TE coefficient), field applied along $a$-axis changes the longitudinal dilation dramatically. These
results are consistent with higher fields study in Ref. \cite{doe02a}. Jumps in $\alpha_a$ and $\alpha_c$
associated with the superconducting transition are small but unequivocal in the data (Fig. \ref{F1}(c)). The
values of $\Delta \alpha_a$ and $\Delta \alpha_c$ at $T_c$ are of the same order of magnitude as in YNi$_2$B$_2$C
\cite{bud06b}, but of the opposite signs. The temperatures associated with the anomalies in the TE coefficients
are consistent with the phase transitions observed in heat capacity (Fig. \ref{F3}).

Using results from the heat capacity (Fig. \ref{F3}) and thermal expansion coefficients (Fig. \ref{F1})
measurements, the uniaxial pressure derivatives of the phase transition temperatures, $T_c$, $T_N$ and $T_{FWM}$,
can be calculated using the thermodynamic Ehrenfest relation for the second order phase transitions:

\begin{displaymath}
\frac {dT_{crit}}{dp_i} = V_{mol} \cdot \Delta \alpha_i(T_{crit}) \cdot \left[ \frac{\Delta
C_p(T_{crit})}{T_{crit}} \right] ^{-1}
\end{displaymath}

\noindent where $V_{mol}$ is a molar volume of the material (for ErNi$_2$B$_2$C, $V_{mol} = 39.2$ cm$^3$) and
$\Delta \alpha_i(T_{crit})$ and $\Delta C_p(T_{crit})$ are the jumps in the $i$-th thermal expansion coefficient
and specific heat at $T_{crit}$. Estimates of the uniaxial pressure derivatives for ErNi$_2$B$_2$C based on the
aforementioned measurements are summarized in Table \ref{tab}. Due to complex behavior of $\alpha_a$ at $T_N$, we
were not able to estimate $dT_N/dp_a$ and $dT_N/dP^*$. The error bars for the uniaxial pressure derivatives listed
in the Table \ref{tab} are rather large, possibly up to $30 - 50$ \% due to signal to noise ratio in the
temperature-dependent TE coefficient data and possible ambiguity in $\Delta T_{crit}$ determination at $T_c$ and
$T_{WFM}$.

In the table $dT_{crit}/dP^* = 2 \cdot dT_{crit}/dp_{ab} + dT_{crit}/dp_c$. The $dT_{crit}/dP^*$ defined in such
way lacks the contribution from the off-diagonal $dT_{crit}/dp_{ij}, i \not= j$ terms that play a role in the
experimentally measured $dT_c/dP$ under hydrostatic pressure. Since the off-diagonal terms are usually
significantly smaller then the diagonal ones, we can still compare the last column of the Table \ref{tab} with the
experimentally measured hydrostatic pressure derivatives. The hydrostatic pressure measurements on ErNi$_2$B$_2$C
are rather sparse: there were two reports on initial ($P \to 0$) $dT_c/dP$ \cite{sch94a,all95a} with the values of
different signs, $-0.82 \times 10^{-2}$ K/kbar and $+1.68 \times 10^{-2}$ respectively, $T_N$ was reported to
increase under pressure \cite{mat00a} and no data for $T_{WFM}$ under pressure are available so far. For $T_c$ our
estimates, within the restrictions mentioned above, are qualitatively consistent with the small and negative
$dT_c/dP$ reported in Ref. \cite{sch94a}. Our results predict high sensitivity of both $T_N$ and $T_{WFM}$ to
uniaxial pressure. Within the above analysis, we can expect that the weak ferromagnetic state is very fragile and
hydrostatic pressure as moderate as 1 kbar will drive $T_{WFM}$ to zero. It will require heat capacity
measurements under pressure to verify this prediction. It is curious that whereas $T_c$ is almost pressure
independent and $T_N$ and $T_{WFM}$ appear to be very pressure sensitive, another perturbation to the material, Co
doping to the Ni site, yields the opposite effect \cite{bud00b}: already 6\% of Co in
Er(Ni$_{1-x}$Co$_x$)$_2$B$_2$C brings $T_c$ down to below 2 K, while 8\% of Co causes decrease of $T_N$ only by
$\sim 0.5$ K and has almost no effect on $T_{WFM}$. Comparison of these two perturbations suggests that steric
effects are well separated from band-filling effects in ErNi$_2$B$_2$C.

\begin{figure}
    \centering
    \includegraphics[height=100mm]{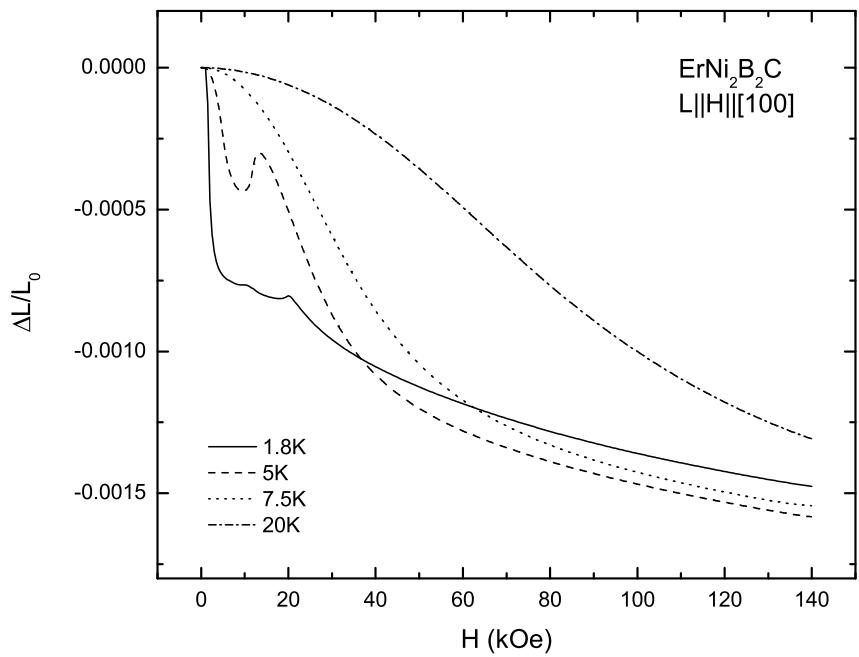}
    \caption{Longitudinal magnetostriction of ErNi$_2$B$_2$C measured for $H \| a$.} \label{F4}
\end{figure}

\begin{figure}
    \centering
    \includegraphics[height=100mm]{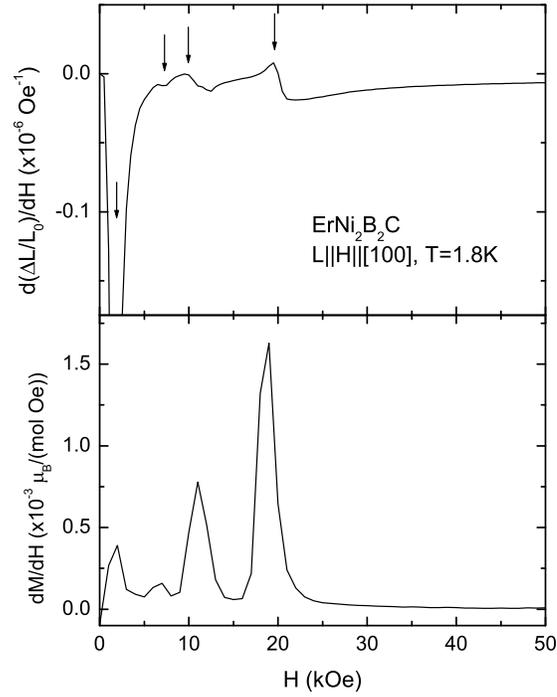}
    \caption{Upper panel - derivative of magnetostriction with respect to magnetic field for $H \leq 50$ kOe.
    Dip at $\sim 1.9$ kOe reaches $-4.75 \times 10^{-7}$ Oe$^{-1}$. Arrows mark features associated with the
    metamagnetic transitions. Lower panel - derivative of magnetization with respect to magnetic field for the
    same orientation.} \label{F5}
\end{figure}

Longitudinal magnetostriction of ErNi$_2$B$_2$C for $L \| H \| a$ is shown in Fig. \ref{F4}. In the overlaping
temperature range our data are consistent with the results of Ref. \cite{doe02a}. At base temperature (1.8 K)
features associated with four metamagnetic transitions can be distinguished in the derivative of magnetostriction
with respect to magnetic field (Fig. \ref{F5}), consistent with (within errors of orientation) with the
magnetization data. The base temperature magnetostriction data suggest that, in agreement with earlier
magnetization studies \cite{can96a,can97a,bud00a} the $H - T$ phase diagram for ErNi$_2$B$_2$C is richer and more
complex than suggested in Ref. \cite{doe02a}.

\textit{In summary}, thermal expansion measurements combined with temperature-dependent specific heat data, allow
us to estimate uniaxial pressure dependencies of three phase transitions in ErNi$_2$B$_2$C, superconducting,
antiferromagnetic and weak ferromagnetic. Our data are roughly consistent with the results of hydrostatic pressure
effect on superconducting transition temperature. Both, N\'eel and weak ferromagnetic ordering temperature are
envisaged to be very sensitive to stress/pressure. Additionally, magnetostriction is shown to be a useful probe
for rich and complex $H - T$ phase diagram in this material.

\ack Ames Laboratory is operated for the U. S. Department of Energy by Iowa State University under Contract No.
W-7405-Eng.-82. This work was supported by the director for Energy Research, Office of Basic Energy Sciences. One
of us (GMS) is supported by the National Science Foundation under DMR-0305397. We thank V. Pelevin for for his
useful suggestion regarding comparative mensuration.

\clearpage

\begin{table}[tbp]
\caption{Changes in specific heat and thermal expansion coefficients at three ordering temperatures, $T_{crit}$,
for superconducting (SC), antiferromagnetic (AFM) and weak ferromagnetic (WFM) phase transitions and estimates of
their anisotropic pressure derivatives.} \label{tab}

\begin{tabular}{lccccccc}

\hline\hline
transition&$T_{crit}$&$\Delta C_p$&$\Delta \alpha_a$&$\Delta \alpha_c$&$dT_{crit}/dp_a$&$dT_{crit}/dp_c$&$dT_{crit}/dP^*$\\

          &(K)&(J/mol K)&($10^{-6}$ K$^{-1}$)&($10^{-6}$ K$^{-1}$)&(K/kbar)&(K/kbar)&(K/kbar)\\
\hline\hline
SC   &10.1&0.24&0.01&-0.08&$1.6 \cdot 10^{-2}$&$-13 \cdot 10^{-2}$&$-9.8 \cdot 10^{-2}$\\
\hline
AFM   &6.1&23.3& &-117&  &$-1.2$&\\
\hline
WFM  &2.3&0.6&-9.0&-1.4&$-1.4$&$-0.2$&$-3$\\
\hline\hline

\end{tabular}
\end{table}

\clearpage

\clearpage

\clearpage

\clearpage

\clearpage

\end{document}